%% file: alice-other-harmonics.tex
\documentclass[aps,prl,twocolumn,superscriptaddress,altaffilletter,nobibnotes,nofootinbib,10pt,showpacs,showkeys,preprintnumbers]{revtex4-1}

\def\pt{\mbox{$p_{\rm t} $}}   
   
\def\snn{\mbox{$\sqrt{s_{_{\rm NN}}}$}}

\newcommand{ \be }{\begin{eqnarray}}
\newcommand{ \ee }{\end{eqnarray}}
\newcommand{ \la }{\langle}
\newcommand{ \ra }{\rangle}

\newcommand{ \eps }{\varepsilon}
\newcommand{ \mean }[1]{\la #1 \ra}

\usepackage{preprintcover}

\PreprintIdNumber{CERN-PH-EP-2011-073}
\PreprintCoverPaperTitle{Higher harmonic anisotropic flow measurements of charged particles in Pb--Pb collisions  
at $\sqrt{s_{_{\rm NN}}} = 2.76$~TeV}
\PreprintCoverAbstract{We report on the first measurement of the triangular $v_3$, quadrangular
$v_4$, and pentagonal $v_5$ charged particle flow in Pb--Pb collisions
at $\snn~=~2.76$~TeV measured with the ALICE detector at the CERN Large Hadron
Collider.  
We show that the triangular flow can be described in terms of 
the initial spatial anisotropy and its fluctuations, which
provides strong constraints on its origin.  
In the most central events, where the elliptic flow $v_2$ and $v_3$
have similar magnitude, a double peaked 
structure in the two-particle azimuthal correlations is observed, 
which is often interpreted as a Mach cone response to fast partons. 
We show that this structure can be naturally explained from the measured anisotropic flow Fourier coefficients.}
\PreprintJournalName{PRL}

\usepackage{graphicx}  
\usepackage{dcolumn}   
\usepackage{bm}        
\usepackage{amssymb}   
\usepackage{lineno}
\usepackage{draftcopy}

\usepackage{color}
\definecolor{dgreen}{cmyk}{1.,0.,1.,0.2}        
\definecolor{orange}{cmyk}{0.,0.353,1.,0.}    

\usepackage[pdftex,bookmarks]{hyperref}

\begin{document}

\title{Higher harmonic anisotropic flow measurements of charged particles in Pb--Pb collisions  
at $\sqrt{s_{_{\rm NN}}} = 2.76$~TeV}

\collaboration{ALICE Collaboration} 
\noaffiliation
\input{authorlist2}

\begin{abstract}
We report on the first measurement of the triangular $v_3$, quadrangular
$v_4$, and pentagonal $v_5$ charged particle flow in Pb--Pb collisions
at $\snn~=~2.76$~TeV measured with the ALICE detector at the CERN Large Hadron
Collider.  
We show that the triangular flow can be described in terms of 
the initial spatial anisotropy and its fluctuations, which
provides strong constraints on its origin.  
In the most central events, where the elliptic flow $v_2$ and $v_3$
have similar magnitude, a double peaked 
structure in the two-particle azimuthal correlations is observed, 
which is often interpreted as a Mach cone response to fast partons. 
We show that this structure can be naturally explained from the measured anisotropic flow Fourier coefficients.

\end{abstract}

\maketitle

The quark-gluon plasma (QGP) is a state of matter whose existence at
high energy density is predicted by Quantum Chromodynamics.  The
creation of this state of matter in the laboratory, and the study of its
properties are the main goals of the ultra-relativistic nuclear collision program.
One of the experimental observables that is sensitive to the
properties of this matter is the azimuthal distribution of particles
in the plane perpendicular to the beam direction. When nuclei collide
at non-zero impact parameter (non-central collisions), the geometrical
overlap region is anisotropic.  This initial spatial asymmetry 
is converted via multiple collisions into an anisotropic momentum
distribution of the produced particles~\cite{Ollitrault:1992bk}.

The azimuthal anisotropy is usually characterized by the Fourier
coefficients~\cite{Voloshin:1994mz,Poskanzer:1998yz}:
\begin{equation}
v_n = \mean{\cos[ n (\phi-\Psi_n)]},
\label{eqFourierCoeff}
\end{equation}
where $\phi$ is the azimuthal angle of the particle, 
$\Psi_n$ is the angle of the initial state spatial plane of symmetry,
and $n$ is the order of the harmonic.  
Because the planes of symmetry $\Psi_n$ are not known experimentally, 
the anisotropic flow coefficients are estimated from measured
correlations between the observed particles.  
The second Fourier coefficient $v_2$ is called elliptic flow and 
has been studied in detail in recent years~\cite{Voloshin:2008dg}. 
Large values of elliptic flow at the LHC were recently observed
by the ALICE collaboration~\cite{Aamodt:2010pa}.

In a non-central heavy ion collision the beam axis and the impact
parameter define the reaction plane $\Psi_{\rm RP}$.  Assuming a
smooth matter distribution in the colliding nuclei, the plane of symmetry is 
the reaction plane $\Psi_n = \Psi_{\rm RP} $ and the odd Fourier coefficients are 
zero by symmetry.    
However due to
fluctuations in the matter distribution, including contributions from fluctuations
in the positions of the participating nucleons in the nuclei, the plane
of symmetry fluctuates event-by-event around the reaction plane.  This
plane of symmetry is determined by the participating nucleons and is
therefore called the participant plane
$\Psi_{\rm PP}$~\cite{Manly:2005zy}.  
Event-by-event fluctuations
of the spatial asymmetry generate additional odd harmonic
symmetry planes $\Psi_{n}$, which are predicted to give rise to the odd harmonics 
like $v_3$ and $v_5$~\cite{Mishra:2007tw,Mishra:2008dm,Takahashi:2009na,Alver:2010gr,
Alver:2010dn,Teaney:2010vd,Luzum:2010sp}.

The large elliptic flow at the Relativistic Heavy Ion
Collider (RHIC)~\cite{Ackermann:2000tr,Adler:2003kt} and at the
LHC~\cite{Aamodt:2010pa} provides compelling evidence for
strongly interacting matter which appears to behave like an almost
perfect (inviscid) fluid~\cite{Kovtun:2004de}.  
Deviations from this ideal case are controlled by  the ratio $\eta/s$ of shear
viscosity to entropy density.  
Because the effect of shear viscosity is to dampen all coefficients, 
with a larger decrease for higher order coefficients~\cite{Teaney:2010vd,Qin:2010pf}, 
it has been argued that the magnitude and transverse momentum dependence of the 
coefficients $v_3$ and $v_5$ is a more sensitive measure
of $\eta/s$~\cite{Alver:2010dn}.  
Therefore a measurement of these
Fourier coefficients at the LHC provides strong constraints on the
initial geometry, its fluctuations, as well as on the shear
viscosity to entropy density ratio.

In this paper we report the first measurement of the anisotropic flow
coefficients $v_3$, $v_4$, and $v_5$ of charged particles in Pb--Pb
collisions at the center of mass energy per nucleon pair
$\snn~=~2.76$~TeV, with the ALICE
detector~\cite{Aamodt:2008zz,Carminati:2004fp,Alessandro:2006yt}.  The
data were recorded in November 2010 during the first run with heavy
ions at the LHC.

For this analysis the ALICE Inner Tracking System (ITS) and the Time
Projection Chamber (TPC) were used to reconstruct charged particle
tracks.  The VZERO counters and the Silicon Pixel Detector (SPD) were
used for the trigger.  The VZERO counters are two scintillator arrays
providing both amplitude and timing information, covering the
pseudorapidity range $2.8 < \eta < 5.1$ (VZERO-A) and $-3.7 < \eta <
-1.7$ (VZERO-C).  The SPD is the innermost part of the ITS, consisting
of two cylindrical layers of hybrid silicon pixel assemblies covering
the range of $|\eta|<2.0$ and $|\eta|<1.4$ for the inner and outer
layer, respectively.  The minimum-bias interaction trigger required 
the following three conditions~\cite{Aamodt:2010cz}:
(i) two pixel chip hits in the outer layer of the silicon pixel
detectors (ii) a signal in VZERO-A (iii) a signal in VZERO-C.  
Deflection of neutral recoils, which is sensitive to the directed
flow of spectators, is measured with two neutron Zero Degree Calorimeters 
(ZDCs) installed on each side, 114 meters  from the interaction point.
Only events with a vertex found within
7~cm  from the center of the detector along the beam line
were used in the analysis. This is to ensure a uniform acceptance in the
central pseudorapidity region $|\eta|<0.8$.  An event sample of
$5 \times 10^6$ Pb--Pb collisions passed the selection criteria
and was  analyzed as a function of collision centrality, 
determined by cuts on the VZERO multiplicity as described
in~\cite{Aamodt:2010cz}.
Based on the strong correlation between the collision centrality determined by the ZDC, TPC, SPD and VZERO 
detectors, the resolution in centrality is found to be $< 0.5$\% RMS for the most central 
collisions (0--5\%), increasing towards 2\% RMS for peripheral collisions (e.g. 70--80\%).
This resolution is also in agreement with our Monte Carlo Glauber~\cite{Miller:2007ri} studies.

The analysis was, as in~\cite{Aamodt:2010pa}, performed using tracks measured 
with only the TPC and for tracks using the ITS and TPC. 
These two measurements have very different acceptance and efficiency corrections, 
and provide an estimate of a possible residual bias which may be present, 
after correction, for small values of the harmonics~\cite{Bilandzic:2010jr}. 
For both measurements, charged particles were selected with high reconstruction efficiency and
minimal contamination from photon conversions and secondary charged
particles produced in the detector material as
described in~\cite{Aamodt:2010pa}.
From Monte Carlo simulations of
{\sc HIJING}~\cite{refHIJING} events using a {\sc GEANT3}~\cite{geant3} detector
simulation and event reconstruction
the estimated contamination is less than 6\% at $p_{\rm t} = 0.2$ GeV/$c$ and drops
below 1\% at $p_{\rm t} > 1$ GeV/$c$. 
In this Letter we present the results obtained using the TPC
standalone tracks, because of the smaller corrections for the azimuthal acceptance.

We report the anisotropic flow coefficients $v_n$ obtained from two-particle correlations and from 
a four-particle cumulant method~\cite{Borghini:2001vi}, 
denoted $v_n\{2\}$ and $v_n\{4\}$, respectively. 
To calculate the four-particle cumulants we used the method proposed
in~\cite{Bilandzic:2010jr}.  The $v_n\{2\}$ and $v_n\{4\}$
measurements have different sensitivity to flow fluctuations and
contributions from nonflow.  The nonflow contribution arises from
correlations between the particles unrelated to the initial geometry.
The contribution from flow fluctuations is positive for $v_n\{2\}$ 
while it is negative for $v_n\{4\}$~\cite{Miller:2003kd}.  
Because the odd
harmonics are expected to be completely due to event-by-event
fluctuations in the initial spatial geometry, the comparison of these two and 
four-particle cumulants provides a strong constraint on the initial spatial 
geometry fluctuations.  

The nonflow contribution to the two-particle correlations is not known
and might be significant.  
We utilize four methods to study and correct for nonflow contributions to the $v_n$\{2\} coefficients.   
First we compare $v_n$\{2\} for like and unlike charge-sign combinations since they have different  contributions from 
resonance decay and jet fragmentation.  
Second we used different pseudorapidity gap requirements between the two particles since larger gaps 
reduce the nonflow contributions.  
Third we utilize HIJING (a pQCD inspired model which does not include flow) to estimate these contributions and, finally 
we estimate the nonflow from the correlations measured in proton--proton collisions.  
All of these methods indicate that nonflow effects are smaller than 10\%.  
In this Letter we use the dependence of the
correlations on pseudorapidity distance between particles as an estimate of nonflow. 

\begin{figure}[thb]
  \begin{center}
    \includegraphics[width=0.48\textwidth]{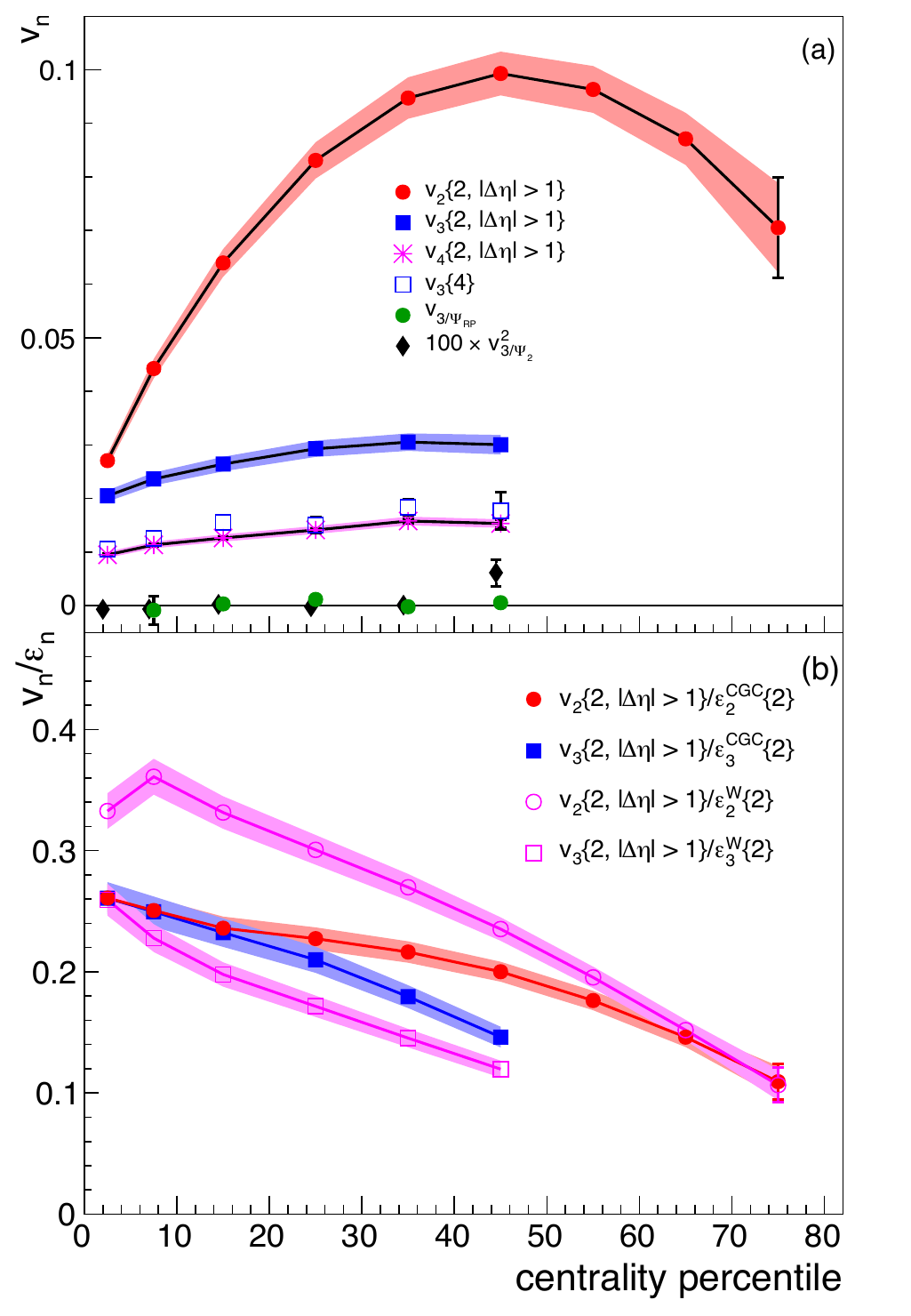}
    \caption{ (color online) a) $v_2$, $v_3$ and $v_4$ integrated over the $\pt$ range
      $0.2 < \pt < 5.0$ GeV/$c$ as a function of event centrality, 
      with the more central (peripheral) collisions shown on the left- (right-)hand side, respectively.
      Full and open squares show $v_3$\{2\} and $v_3$\{4\}, respectively. 
      In addition we show $v^2_{3/\Psi_2}$ and $v_{3/\Psi_{\rm RP}}$,
       which represent the 
        triangular flow measured relative to the second order event plane and
        the reaction plane, respectively (for the definitions, see text).  
       b) $v_2$\{2, $|\Delta\eta| > 1$\} and $v_3$\{2, $|\Delta\eta| > 1$\} divided by the
      corresponding eccentricity versus centrality percentile for Glauber~\cite{Miller:2007ri} and MC-KLN CGC~\cite{Drescher:2007ax} 
      initial conditions.  }
    \label{v3_int_flow}
  \end{center}
\end{figure} 

Figure~\ref{v3_int_flow}a shows $v_2$, $v_3$ and $v_4$ integrated over the $\pt$ range
$0.2 < \pt < 5.0$ GeV/$c$ as a function of centrality.  The $v_2$\{2\}, $v_3$\{2\}  and $v_4$\{2\}
are shown for particles with $|\Delta\eta| > 1.0$ and corrected for the estimated remaining 
nonflow contribution based on the correlation measured in HIJING. 
The total systematic uncertainty is shown as a band and fully includes this residual correction.
The measured $v_3$ is smaller than $v_2$ and does not depend strongly on centrality. 
The $v_3$ is compatible with predictions for Pb--Pb 
collisions from a hydrodynamic model calculation with Glauber initial 
conditions and $\eta/s =0.08$ and larger than for MC-KLN CGC initial conditions with 
$\eta/s=0.16$~\cite{Alver:2010dn}, suggesting a small value of  
$\eta/s$ for the matter created in these collisions. 
The $v_3$\{4\} is about a factor two smaller than the two-particle measurement 
which can, as explained in~\cite{Bhalerao:2011yg}, 
be understood if $v_3$ originates predominantly from event-by-event
fluctuations of the initial spatial geometry.  
For these event-by-event fluctuations of the spatial geometry, 
the symmetry plane $\Psi_3$ is expected to be  
uncorrelated (or correlated very weakly~\cite{Nagle:2010zk}) with the 
reaction plane $\Psi_{\rm RP}$, and with $\Psi_2$.  
We evaluate the correlations between
$\Psi_3$ and $\Psi_{\rm RP}$ using the first-order event plane
  from the ZDC via $v_{3/\Psi_{\rm RP}}=\left< \cos (3\phi_1
-3\Psi_{\rm RP})\right>$ and the correlation between $\Psi_3$ and
$\Psi_{2}$ with a five-particle correlator $\left< \cos (3\phi_1 +
3\phi_2 - 2\phi_3 - 2\phi_4 - 2\phi_5)\right>/v_{2}^{3}=v_{3/\Psi_2}^{2}$.
In Fig.~\ref{v3_int_flow}a $v_{3/\Psi_{\rm RP}}$ and 
$v_{3/\Psi_{2}}^2$ are shown as a function of
centrality. These correlations are
indeed, within uncertainties, consistent with zero as expected from a  
triangular flow that originates predominantly from event-by-event
fluctuations of the initial spatial geometry.   

To investigate the role of viscosity further we calculate the ratios $v_2/\varepsilon_2$ 
and $v_3/\varepsilon_3$, where $\varepsilon_2$
and $\varepsilon_3$ are the ellipticity and triangularity of the initial spatial geometry, 
defined by:
\begin{equation}
\varepsilon_n = - \frac{\left< r^2 \cos n(\phi-\Psi_{n})\right>}{\left< r^2\right>} 
\label{eq_eccentricities}
\end{equation}
where the brackets denote an average which traditionally is
taken over the position of participating (wounded) nucleons in a Glauber model~\cite{Miller:2007ri}.

Under the assumption that $v_n$ is proportional to
$\varepsilon_n$, $v_n$\{2\} is
proportional to $\varepsilon_n$\{2\}~\cite{Miller:2003kd}.  
Figure ~\ref{v3_int_flow}b shows the ratios $v_n/\varepsilon_n$ for 
eccentricities calculated with a Glauber and a MC-KLN CGC~\cite{Drescher:2007ax} model,
denoted by $\varepsilon_n^W$\{2\} and $\varepsilon_n^{\rm CGC}$\{2\}, respectively.
We find that for a Glauber model the magnitude of
$v_3\{2\}/\varepsilon_3$\{2\}  is smaller than $v_2\{2\}/\varepsilon_2$\{2\}, which would indicate significant viscous corrections.
For MC-KLN CGC calculations the ratios $v_2\{2\}/\varepsilon_2$\{2\}  and $v_3\{2\}/\varepsilon_3$\{2\} 
are almost equal for the most central collisions, as expected for an almost ideal fluid~\cite{Alver:2010dn}. 
In addition, we notice that the ratio 
$v_3\{2\}/\varepsilon_3$\{2\} decreases faster than $v_2\{2\}/\varepsilon_2\{2\}$ toward more peripheral 
collisions which is expected due to larger viscous corrections to $v_3$.

The centrality dependence of the triangular flow differs 
significantly from that of elliptic flow. 
This might be due to two reasons: either the centrality dependence of the spatial ellipticity and 
triangularity are different, and/or the viscous effects are different.  
However, in a small centrality range, such as 0--5\%, viscous effects do not change much 
and there one might be directly sensitive to the change in the initial spatial geometry.
Our calculations show that even in this small centrality range,
the ratio $\eps_2/\eps_3$ changes significantly which allows
us to investigate further the geometrical origin of elliptical and triangular flow.
\begin{figure}[thb]
  \begin{center}
    \includegraphics[width=0.48\textwidth]{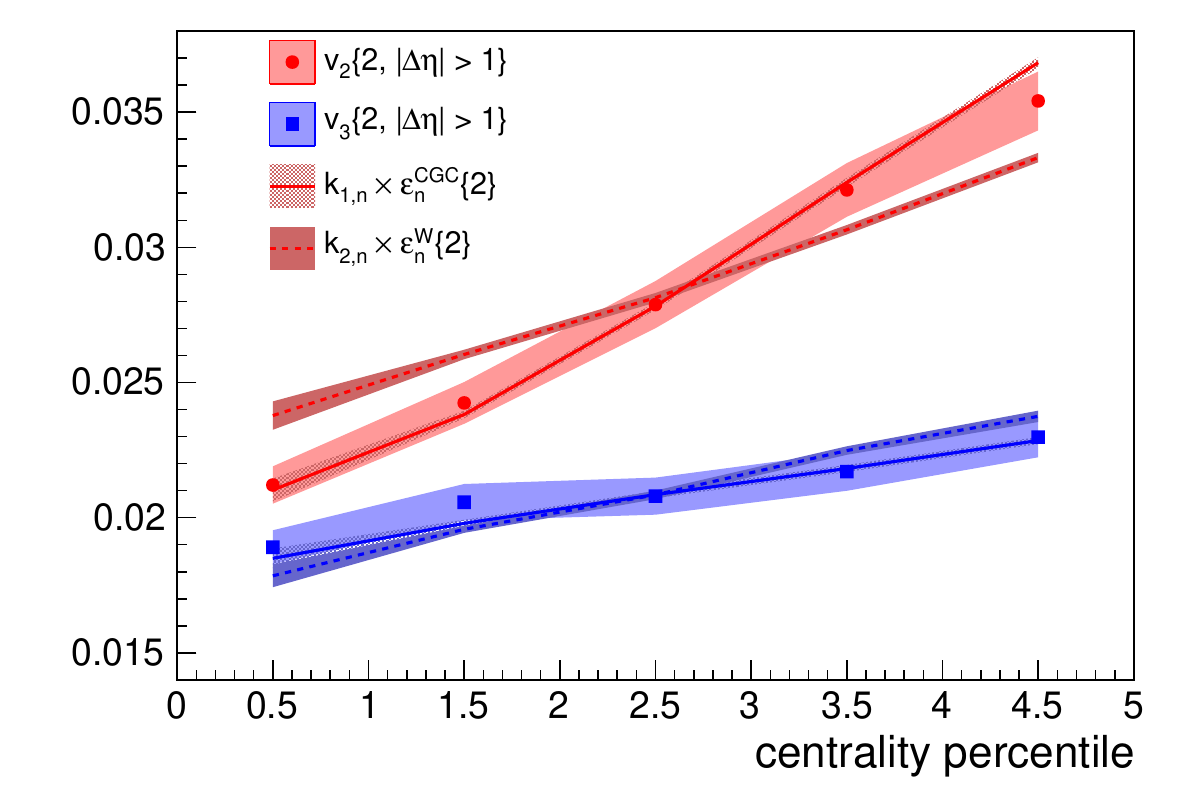}  
    \caption{ (color online) $v_2$ and $v_3$ as a function of centrality
      for the 5\% most central collisions compared to calculations
      of the spatial eccentricities, $\varepsilon_{n}^W$\{2\} and $\varepsilon_{n}^{\rm CGC}$\{2\}.
      The eccentricities have been scaled to match the 2-3\% data using $k_1$ and $k_2$.
      }
    \label{v2_v3_epsilon_very_central}
  \end{center}
\end{figure} 
In Fig.~\ref{v2_v3_epsilon_very_central} $v_2$\{2\} and $v_3$\{2\} are
plotted in 1\% centrality bins for the 5\% most central collisions. 
We observe that $v_3$\{2\} does not change much versus centrality (as
would be expected if $v_3$ is dominated by event-by-event fluctuations
of the initial geometry) while, $v_2$\{2\}
increases by about 60\%.  We compare this dependence of $v_n$\{2\} to
the centrality dependence of the eccentricities $\varepsilon_n$\{2\}
for initial conditions from MC-KLN CGC and Monte Carlo Glauber.  
We observe that the weak dependence of $v_3$\{2\} is described by both calculations 
while the relative strong dependence of $v_2$\{2\} on centrality is only described for the 
MC-KLN CGC initial conditions.

\begin{figure}[thb]
  \begin{center}
    \includegraphics[width=0.46\textwidth]{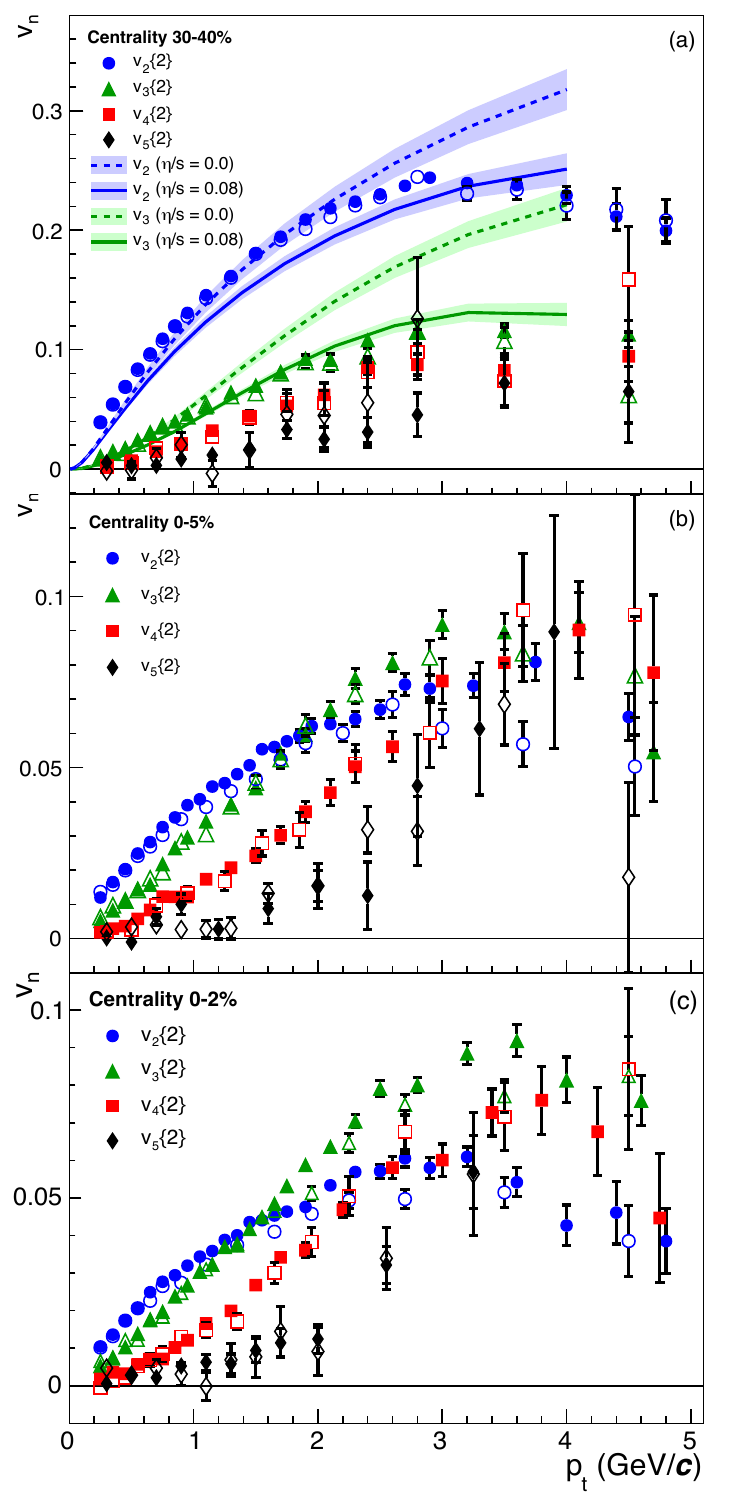}
    \caption{ (color online) $v_2$, $v_3$, $v_4$, $v_5$ as a function of
    transverse momentum and for three event centralities. The full, open symbols 
    are for  $\Delta\eta > 0.2$ and  $\Delta\eta > 1.0$, respectively. 
    a) 30\%--40\% compared to hydrodynamic model calculations
    b) 0--5\% centrality percentile
    c) 0--2\% centrality percentile.
    }
    \label{harmonics}
  \end{center}
\end{figure} 
The harmonics $v_2$\{2\}, $v_3$\{2\}, $v_4$\{2\} and $v_5$\{2\} as a 
function of transverse momentum are shown for the 30\%--40\%, 0--5\%,
and 0--2\% centrality classes in Fig.~\ref{harmonics}.
For the
30\%--40\% centrality class the results are compared to hydrodynamic
predictions using Glauber initial conditions for different values of $\eta/s$~\cite{Schenke:2011tv}.  
We observe that, at low-$\pt$, the
different $\pt$-dependence of $v_2$ and $v_3$ is
described well by these hydrodynamic predictions.  However, the
magnitude of $v_2$($\pt$) is better described by $\eta/s=0$ while for
$v_3$($\pt$) $\eta/s=0.08$ provides a better description.  
We anticipate future comparisons utilizing MC-KLN initial conditions. 

For central collisions 0-5\% we
observe that at $\pt \approx$ 2 GeV/$c$ $v_3$ becomes equal to $v_2$ and
at $\pt \approx$ 3 GeV/$c$ $v_4$ also reaches the same magnitude as $v_2$
and $v_3$.  For more central collisions 0-2\%, we observe that $v_3$
becomes equal to $v_2$ at lower $\pt$ and reaches significantly larger
values than $v_2$ at higher-$\pt$. The same is true for $v_4$ compared
to $v_2$.  

We compare the structures found with azimuthal correlations between
triggered and associated particles to those described by the measured 
$v_n$ components. The two-particle azimuthal correlations are measured by
calculating: 
\begin{equation}
C(\Delta\phi) \equiv \frac{N_{\rm mixed}}{N_{\rm same}} \frac{{\rm d}N_{\rm same}/{\rm d}\Delta\phi}{{\rm d}N_{\rm mixed}/{\rm d}\Delta\phi} ,
\end{equation}
where $\Delta\phi =\phi_{trig} -\phi_{assoc}$. d$N_{\rm same}$/d$\Delta\phi$ (d$N_{\rm mixed}$/d$\Delta\phi$) is the 
number of associated particles as function of $\Delta\phi$ within the same (different) event, 
and $N_{\rm same}$ ($N_{\rm mixed}$) the total number of associated particles in d$N_{\rm same}$/d$\Delta\phi$ (d$N_{\rm mixed}$/d$\Delta\phi$).
\begin{figure}[thb]
  \begin{center}
    \includegraphics[width=0.47\textwidth]{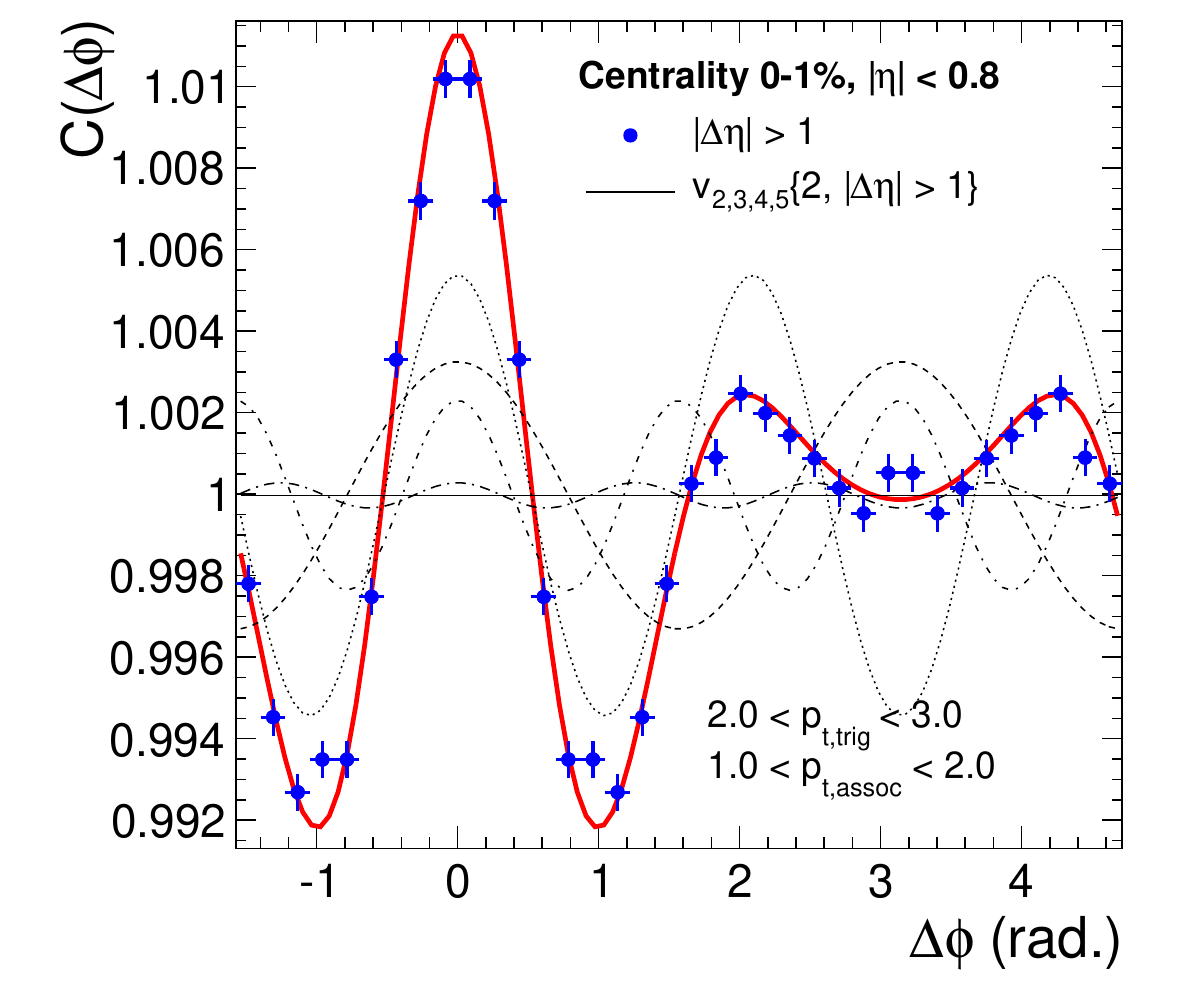}
    \caption{ (color online) The two-particle azimuthal correlation, 
    measured in $0 < \Delta\phi < \pi$ and shown symmetrized over $2\pi$, between a trigger particle with $2 < \pt < 3$ GeV/$c$ and an
    associated particle with $1 < \pt < 2$ GeV/$c$ for the 0--1\%
    centrality class. The solid red line shows the sum of the measured 
    anisotropic flow Fourier coefficients $v_2$, $v_3$, $v_4$ and $v_5$ (dashed lines).}
    \label{twoparticle}
  \end{center}
\end{figure} 
Figure~\ref{twoparticle} shows the azimuthal correlation
observed in very central collisions 0--1\%, for trigger particles in
the range $2 < \pt < 3$ GeV/$c$ with
associated particles in $1 < \pt < 2$ GeV/$c$ for pairs in $|\Delta\eta| > 1$.  
We observe a clear doubly-peaked correlation structure centered
opposite to the trigger particle.  This feature has been observed at
lower energies in broader centrality bins~\cite{:2008cqb,Aggarwal:2010rf}, but only after
subtraction of the elliptic flow component.
This two-peak structure has been interpreted as an indication for
various jet-medium modifications (i.e. Mach cones)~\cite{:2008cqb,Aggarwal:2010rf} and more recently as 
a manifestation of triangular flow~\cite{Alver:2010gr,Alver:2010dn,Teaney:2010vd,Luzum:2010sp}.
We therefore compare the azimuthal correlation shape expected from $v_2$, $v_3$, $v_4$ 
and $v_5$ evaluated at corresponding transverse momenta 
with the measured two-particle azimuthal triggered correlation and find that
the combination of these harmonics gives a natural description of the
observed correlation structure on the away-side.

In summary, we have presented the first measurement at the LHC of
triangular $v_3$, quadrangular $v_4$ and pentagonal particle flow
$v_5$.  We have shown that the triangular flow and its fluctuations
can be understood from the initial spatial anisotropy. 
The transverse momentum dependence of $v_2$ and $v_3$ compared to
model calculations favors a small value of the shear viscosity to
entropy ratio $\eta/s$.  For the 5\% most central collisions we have
shown that $v_2$ rises strongly with centrality
in 1\% centrality percentiles.  The strong change in $v_2$
and the small change in $v_3$ as a function of centrality in these 1\%
centrality percentile classes follow the
centrality dependence of the corresponding spatial anisotropies. 
The two-particle azimuthal correlation for the 0--1\% centrality
class exhibits a double peak structure around $\Delta\phi \sim \pi$ 
(the ``away side") without the subtraction of elliptic flow.
We have shown that the
measured anisotropic flow Fourier coefficients give a
natural description of this structure.

\section{acknowledgements}
\input{acknowledgements.tex}

\end{document}

%% file: authorlist2.tex
\author{K.~Aamodt}
\altaffiliation{}
\affiliation{Department of Physics and Technology, University of Bergen, Bergen, Norway}

\author{B.~Abelev}
\altaffiliation{}
\affiliation{Lawrence Livermore National Laboratory, Livermore, California, United States}

\author{A.~Abrahantes~Quintana}
\altaffiliation{}
\affiliation{Centro de Aplicaciones Tecnol\'{o}gicas y Desarrollo Nuclear (CEADEN), Havana, Cuba}

\author{D.~Adamov\'{a}}
\altaffiliation{}
\affiliation{Nuclear Physics Institute, Academy of Sciences of the Czech Republic, \v{R}e\v{z} u Prahy, Czech Republic}

\author{A.M.~Adare}
\altaffiliation{}
\affiliation{Yale University, New Haven, Connecticut, United States}

\author{M.M.~Aggarwal}
\altaffiliation{}
\affiliation{Physics Department, Panjab University, Chandigarh, India}

\author{G.~Aglieri~Rinella}
\altaffiliation{}
\affiliation{European Organization for Nuclear Research (CERN), Geneva, Switzerland}

\author{A.G.~Agocs}
\altaffiliation{}
\affiliation{KFKI Research Institute for Particle and Nuclear Physics, Hungarian Academy of Sciences, Budapest, Hungary}

\author{A.~Agostinelli}
\altaffiliation{}
\affiliation{Dipartimento di Fisica dell'Universit\`{a} and Sezione INFN, Bologna, Italy}

\author{S.~Aguilar~Salazar}
\altaffiliation{}
\affiliation{Instituto de F\'{\i}sica, Universidad Nacional Aut\'{o}noma de M\'{e}xico, Mexico City, Mexico}

% ...

%% file: acknowledgements.tex
The ALICE collaboration would like to thank all its engineers and technicians for their invaluable contributions to the construction of the experiment and the CERN accelerator teams for the outstanding performance of the LHC complex.
The ALICE collaboration acknowledges the following funding agencies for their support in building and
running the ALICE detector:
Calouste Gulbenkian Foundation from Lisbon and Swiss Fonds Kidagan, Armenia;
Conselho Nacional de Desenvolvimento Cient\'{\i}fico e Tecnol\'{o}gico (CNPq), Financiadora de Estudos e Projetos (FINEP),
Funda\c{c}\~{a}o de Amparo \`{a} Pesquisa do Estado de S\~{a}o Paulo (FAPESP);
National Natural Science Foundation of China (NSFC), the Chinese Ministry of Education (CMOE)
and the Ministry of Science and Technology of China (MSTC);
Ministry of Education and Youth of the Czech Republic;
Danish Natural Science Research Council, the Carlsberg Foundation and the Danish National Research Foundation;
The European Research Council under the European Community's Seventh Framework Programme;
Helsinki Institute of Physics and the Academy of Finland;
French CNRS-IN2P3, the `Region Pays de Loire', `Region Alsace', `Region Auvergne' and CEA, France;
German BMBF and the Helmholtz Association;
Hungarian OTKA and National Office for Research and Technology (NKTH);
Department of Atomic Energy and Department of Science and Technology of the Government of India;
Istituto Nazionale di Fisica Nucleare (INFN) of Italy;
MEXT Grant-in-Aid for Specially Promoted Research, Ja\-pan;
Joint Institute for Nuclear Research, Dubna;
 %
National Research Foundation of Korea (NRF);
CONACYT, DGAPA, M\'{e}xico, ALFA-EC and the HELEN Program (High-Energy physics Latin-American--European Network);
Stichting voor Fundamenteel Onderzoek der Materie (FOM) and the Nederlandse Organisatie voor Wetenschappelijk Onderzoek (NWO), Netherlands;
Research Council of Norway (NFR);
Polish Ministry of Science and Higher Education;
National Authority for Scientific Research - NASR (Autoritatea Na\c{t}ional\u{a} pentru Cercetare \c{S}tiin\c{t}ific\u{a} - ANCS);
Federal Agency of Science of the Ministry of Education and Science of Russian Federation, International Science and
Technology Center, Russian Academy of Sciences, Russian Federal Agency of Atomic Energy, Russian Federal Agency for Science and Innovations and CERN-INTAS;
Ministry of Education of Slovakia;
CIEMAT, EELA, Ministerio de Educaci\'{o}n y Ciencia of Spain, Xunta de Galicia (Conseller\'{\i}a de Educaci\'{o}n),
CEA\-DEN, Cubaenerg\'{\i}a, Cuba, and IAEA (International Atomic Energy Agency);
The Ministry of Science and Technology and the National Research Foundation (NRF), South Africa;
Swedish Reseach Council (VR) and Knut $\&$ Alice Wallenberg Foundation (KAW);
Ukraine Ministry of Education and Science;
United Kingdom Science and Technology Facilities Council (STFC);
The United States Department of Energy, the United States National
Science Foundation, the State of Texas, and the State of Ohio.